\documentclass[]{aastex631}

\shorttitle{Crescent-Shaped Molecular Outflow}
\shortauthors{Harada et al.}

\usepackage[whole]{bxcjkjatype}
\usepackage{amsmath}
\usepackage{amssymb}
\newcommand{\msun}{\thinspace M_\odot}

\newcommand{\kms}{\,km\,s$^{-1}$}

\begin{document}

\title{Crescent-Shaped Molecular Outflow from the Intermediate-mass Protostar DK~Cha Revealed by ALMA}

\correspondingauthor{Naoto Harada}
\email{harada.naoto.450@s.kyushu-u.ac.jp}

\author[0000-0002-8217-7509]{Naoto Harada}
\affiliation{Department of Earth and Planetary Sciences, Faculty of Science, Kyushu University, 744 Motooka, Nishi-ku, Fukuoka 819-0395, Japan}

\author[0000-0002-2062-1600]{Kazuki Tokuda}
\affiliation{Department of Earth and Planetary Sciences, Faculty of Science, Kyushu University, 744 Motooka, Nishi-ku, Fukuoka 819-0395, Japan}
\affiliation{National Astronomical Observatory of Japan, National Institutes of Natural Sciences, 2-21-1 Osawa, Mitaka, Tokyo 181-8588, Japan}
\affiliation{Department of Physics, Graduate School of Science, Osaka Metropolitan University, 1-1 Gakuen-cho, Naka-ku, Sakai, Osaka 599-8531, Japan}

\author{Hayao Yamasaki}
\affiliation{Department of Earth and Planetary Sciences, Faculty of Science, Kyushu University, 744 Motooka, Nishi-ku, Fukuoka 819-0395, Japan}

\author[0000-0001-5817-6250]{Asako Sato}
\affiliation{Department of Earth and Planetary Sciences, Faculty of Science, Kyushu University, 744 Motooka, Nishi-ku, Fukuoka 819-0395, Japan}

\author{Mitsuki Omura}
\affiliation{Department of Earth and Planetary Sciences, Faculty of Science, Kyushu University, 744 Motooka, Nishi-ku, Fukuoka 819-0395, Japan}

\author[0000-0002-4317-767X]{Shingo Hirano}
\affiliation{Department of Astronomy, School of Science, University of Tokyo, Tokyo 113-0033, Japan}

\author[0000-0001-7826-3837]{Toshikazu Onishi}
\affiliation{Department of Physics, Graduate School of Science, Osaka Metropolitan University, 1-1 Gakuen-cho, Naka-ku, Sakai, Osaka 599-8531, Japan}

\author[0000-0002-1411-5410]{Kengo Tachihara}
\affiliation{Department of Physics, Nagoya University, Furo-cho, Chikusa-ku, Nagoya 464-8601, Japan}

\author[0000-0002-0963-0872]{Masahiro N. Machida}
\affiliation{Department of Earth and Planetary Sciences, Faculty of Science, Kyushu University, 744 Motooka, Nishi-ku, Fukuoka 819-0395, Japan}

\begin{abstract}
We report on an Atacama Large Millimeter/submillimeter Array (ALMA) study of the Class~I or II intermediate-mass protostar DK~Cha in the Chamaeleon~II region.
The $^{12}$CO\,($J$ = 2--1) images have an angular resolution of $\sim 1''$ ($\sim 250$\,au) and show high-velocity blueshifted ($\gtrsim$70\kms) and redshifted ($\gtrsim$50\kms) emissions which have 3000\,au scale crescent-shaped structures around the protostellar disk traced in the 1.3\,mm continuum.
Because the high-velocity components of the CO emission are associated with the protostar, we concluded that the emission traces the pole-on outflow.
The blueshifted outflow lobe has a clear layered velocity gradient with a higher velocity component located on the inner side of the crescent shape, which can be explained by a model of an outflow with a higher velocity in the inner radii.
Based on the directly driven outflow scenario, we estimated the driving radii from the observed outflow velocities and found that the driving region extends over two orders of magnitude.
The $^{13}$CO emission traces a complex envelope structure with arc-like substructures with lengths of $\sim$1000\,au.
We identified the arc-like structures as streamers because they appear to be connected  to a rotating infalling envelope.
DK~Cha is useful for understanding characteristics that are visible by looking at nearly face-on configurations of young protostellar systems, providing an alternative perspective for studying the star-formation process.
\end{abstract}

\keywords{Star formation (1569) --- Protostars (1302) --- Circumstellar envelopes(237) --- Circumstellar disks(235) --- Molecular clouds (1072)}

\section{Introduction} \label{sec:intro}

Protostars and their surrounding disks are formed through the gravitational collapse of dense molecular cloud cores.
Protostars eventually evolve into main-sequence stars via the main accretion phase, which is the most dynamic stage in the star formation process due to the presence of active gas inflow and ejection.
The balance between the two processes (gas inflow and ejection) directly determines the mass of protostars and disks, and thus observational investigations of the inflow/outflow motions are fundamental subjects in star formation studies.
Because molecular outflows play the most significant role in transporting both the gas mass and a large amount of angular momentum originating from the parental molecular cloud cores,
it is important to observe the detailed structure of outflows to verify star formation theory.

Molecular outflows are universally observed around protostars across the Local Group of galaxies (\citealt{Tokuda2022} and references therein), and their physical characteristics, such as momentum and force, are diverse and somewhat depend on the protostellar luminosity \citep[e.g.,][]{bontemps1996,Beuther2002,wu2004}.
Outflows interact with the surrounding gas from the parental core scale to the molecular cloud scale, and their impact on star-forming regions has been discussed \citep[e.g.,][]{lada1985,fukui1993,matzner2000,maclow2004,arce2007,pudritz2019}.
According to recent high-resolution millimeter and sub-millimeter observations toward the solar neighborhood targets, a common feature is relatively low-velocity flows with a wide opening angle.
In addition, a few sources exhibit high-velocity ($\sim$100\,km\,s$^{-1}$) collimated flows \citep[e.g.,][]{bachiller1996,hirano2010,matsushita2019}.
Magnetohydrodynamic (MHD) simulations predict that, unlike the classical  jet-entrainment scenario, a magnetocentrifugal force drives the outflow directly from the rotating circumstellar disk for the whole range of protostellar masses  \citep[e.g.,][]{machida2013,matsushita2017,commercon2022}.
Some observations have reported strong evidence for the directly driven scenario \citep[e.g.,][]{bjerkeli2016,hirota2017,matsushita2019}.
Although these observations provide excellent evidence to constrain the outflow driving mechanism, their projected characteristics in the plane of the sky are highly sensitive to various factors, such as the evolutionary stage, the initial conditions of gravitational collapse \citep[e.g.,][]{hirano2020,machida2020,tsukamoto2020}, and the inclination angles for the targets.
Hence, a case-study characterization toward specific targets is still needed to further understand the outflow phenomena in the main accretion phase.

There is a lack of detailed studies observing outflows driven from protostellar systems with small inclination angles (hereafter, pole-on outflows).
Thus, it is crucial to observe nearly face-on systems because this allows us to look into regions close to the protostars \citep[e.g.,][]{motogi2013,motogi2016,motogi2017,motogi2019,fernandez_lopez2020}.
More massive protostars should be suitable targets for such observations because they have larger disks and more powerful outflows \citep[e.g.,][]{wu2004}. 
In particular, outflows appearing around more massive stars are larger and have a fairly wide range of velocity \citep[e.g.,][]{matsushita2017}.
However, there are only a few high-mass stars within a few hundred parsecs of the Sun.
It is thus reasonable to observe intermediate-mass protostellar systems in the solar neighborhood to spatially resolve pole-on outflows.

Recently, highly complex gas structures around protostars have been revealed by Atacama Large Millimeter/submillimeter Array (ALMA) high-resolution studies.
Molecular line observations reported asymmetric, arc-like gas structures around low-mass protostellar sources \citep[e.g.,][]{tokuda2014,tokuda2018} owing to complex gas dynamics.
Although the origin of such peculiar structures is still under debate and depends on source to source, one of the interpretations is a non-axisymmetric accretion flow onto protostars, called $``$streamers$"$ \citep[][and references therein]{pineda2022}.
\cite{okoda2021} discovered a secondary outflow with a direction perpendicular to the main outflow around a single Class~0 protostar IRAS~15398-3359.
These observations indicate that a classical picture does not apply to some protostellar systems.
Thus, in addition to investigating outflow activities, characterizing complex envelope structures can also be an important approach to exploring the diversity of star formation processes.

IRAS~12496-7650, also known as the Herbig Ae star DK~Cha, is the most luminous IRAS source in the Chamaeleon~II star-forming region \citep{hughes1989,hughes1991}.
Early observations discovered a molecular outflow associated with DK~Cha.
A single-dish study by \citet{knee1992} reported a very high-velocity blueshifted ($|v_{\rm LSR}-v_{\rm sys}| \sim 70$\kms) $^{12}$CO\,($J$ = 1--0) emission, which is most likely an outflow component from protostellar objects.
\citet{van_kempen2009} and \citet{yildiz2015} revealed an overlap of outflow between blueshifted and redshifted lobes by the Atacama Pathfinder EXperiment (APEX) observations with a beam size of $\gtrsim 8''$, and concluded that the system is in a nearly face-on configuration.
However, the detailed structure of the outflow is unclear due to the lack of angular resolution in previous studies.
As described in Sections \ref{sec:obs}--\ref{sec:discussion}, we attempted to spatially resolve the outflow based on ALMA archival data for DK~Cha, but the high-resolution view provided peculiar and complex gas structures with an arc or crescent-shaped morphology in the high-velocity emission that forced us to reconsider the simple outflow interpretation.

In this paper, we report the results of our analysis of archival data from ALMA observations toward DK~Cha.
Section~\ref{sec:obs} describes the ALMA observations and data reduction.
In Section~\ref{sec:results}, we present the results for continuum, $^{12}$CO\,($J$ = 2--1) and $^{13}$CO\,($J$ = 2--1) emissions.
We discuss the interpretation of the oddly shaped outflow revealed by $^{12}$CO and the arc-like structures revealed by $^{13}$CO in Section~\ref{sec:discussion}.
Our conclusions are presented in Section~\ref{sec:conclusions}.

\section{Observations and Target descriptions} \label{sec:obs}

We used the ALMA archival data (project ID: 2019.1.01792.S), which was obtained in Cycle~7 in Band~6 (230\,GHz, 1.3\,mm).
The project targeted 125 protostars in nearby star-forming regions.
The observations toward our target DK Cha centered at ($\alpha_{\rm ICRS}$, $\delta_{\rm ICRS}$) = (12$^{\rm h}$53$^{\rm m}$17$^{\rm s}$.20, $-77^\circ 07'$10\farcs00) were carried out on 2019 October 30 and November 5, using the ALMA main array in its C43-3 configuration.
Two spectral windows targeted $^{12}$CO and $^{13}$CO\,($J$ = 2--1) with a frequency resolution of 61.035\,kHz
and bandwidths of 117\,MHz (1920 channels) and 59\,MHz (960 channels), respectively.
Another spectral window was used for continuum observations with an aggregated bandwidth of 1.875\,GHz.

The data were processed with the Common Astronomy Software Application (CASA) package \citep{casa2022} version 5.6.1-8.
We used the \textit{``tclean''} task in the imaging process with the \textit{``multi-scale''} deconvolver to recover the extended emission.
We applied the \textit{``natural''} weighting with an image grid of 0\farcs13 and velocity resolutions of 0.5\kms.
The synthesized beam size for both the continuum and $^{12}$CO was $\sim 1\farcs1 \times 0\farcs8 \simeq 270\,{\rm au} \times 200\,{\rm au}$ at the distance of DK~Cha \citep[$243.7\pm22.0$\,pc,][]{gaia2021}.
The (1$\sigma$) rms of the molecular line emission, $^{12}$CO\,($J$ = 2--1), and the continuum were 0.26\,K at velocity resolutions of 0.5\kms and 2.1\,mJy beam$^{-1}$, respectively.

DK~Cha has a protostellar mass of 1.4--2$\msun$ \citep{spezzi2008,villenave2021}, a bolometric luminosity of 24--35.4\,$L_\odot$ \citep{spezzi2008,van_kempen2010,kristensen2012,yang2018}, a bolometric temperature of 569--592.0\,K \citep{kristensen2012,yang2018}, and a systemic velocity $v_{\rm sys}$ of 2--4\kms \citep{van_kempen2009,kristensen2012,yildiz2013}.
The evolutionary stage of this object is categorized as Class~I or II \citep{alcala2008,wampfler2013,yang2018}.

\section{Results} \label{sec:results}

\subsection{Continuum Source and $^{12}$CO High-velocity Emission} \label{sec:12co}

\begin{figure*}[htbp]
\centering
\includegraphics[width=180mm]{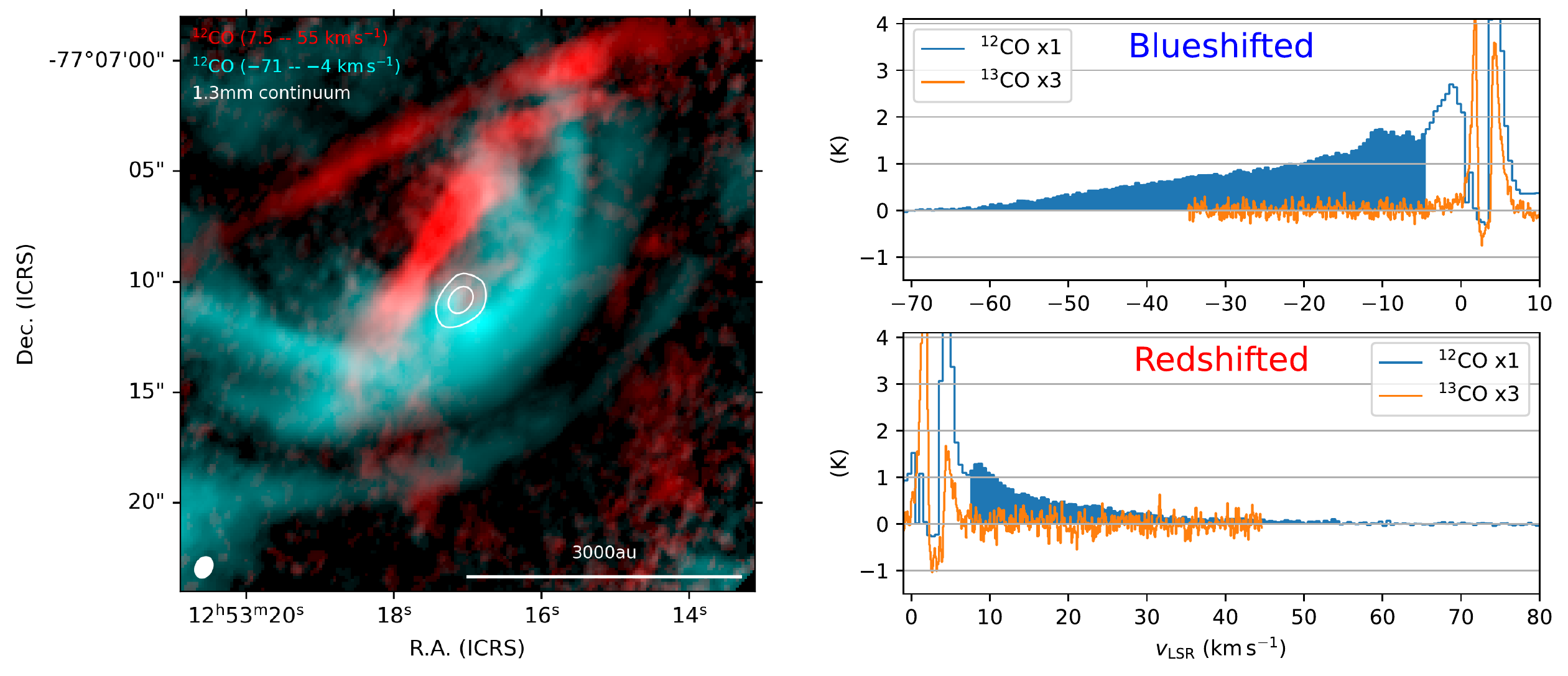}
\caption{
Structure of DK Cha (IRAS 12496-7650) outflow revealed by ALMA.
(Left) Blueshifted and redshifted $^{12}$CO integrated intensity distributions plotted with the 1.3\,mm continuum emission.
The cyan and red colors show the distributions of the blueshifted ($-71$--$-4$\kms) and redshifted (7.5--55\kms) components, respectively.
We used only emissions above $3\sigma$ for each pixel.
The systemic velocity is $\simeq 3$\kms, which was determined by single-dish C$^{18}$O observations \ \citep[e.g.,][]{yildiz2013}.
The white contours show the 1.3\,mm continuum emission with 10 and $100 \sigma$, where $1\sigma$ is the 2.1\,mJy beam$^{-1}$.
The white ellipse at the lower left corner indicates the ALMA synthesized beam; $1\farcs06 \times 0\farcs82$ with a position angle of $-27^\circ$.
(Right) Average spectra of the $>3 \sigma$ blueshifted (top) and redshifted (bottom) components.
The blue and orange lines represent the spectra of $^{12}$CO and $^{13}$CO, respectively.
The $^{13}$CO intensity is multiplied by a factor of 3 for visualization purposes.
The blue-filled areas indicate the integrated velocity ranges.
\label{fig:rgb}}
\end{figure*}

We show the 1.3\,mm continuum and $^{12}$CO\,($J$ = 2--1) emission in this subsection.
The white contours in the left panel of Figure~\ref{fig:rgb} show the 1.3\,mm continuum emission.
The emitting region is compact, comparable to the beam size.
A 2D Gaussian fit to the image gave a deconvolved size of ($503\pm23$\,mas)$\times$($306\pm22$\,mas) and a position angle of $128.8\pm5.3^\circ$.
The total flux is $562.4\pm5.4$\,mJy.
These values are consistent with those reported by \citet{villenave2021}, who used other ALMA archival data with a higher angular resolution of $0\farcs51\times0\farcs28$.
The cyan and red colors represent the blueshifted and redshifted $^{12}$CO\,($J$ = 2--1) integrated intensity distributions, respectively.
The right panels of Figure \ref{fig:rgb} show the average spectra of blueshifted (top) and redshifted (bottom) components.
We defined the $^{12}$CO integration ranges of the maps as follows; the high-velocity sides are the velocity at which the $^{12}$CO emission is prominently detected judging from the averaged profiles, and we excluded the lower-velocity range around the systemic velocity of $\sim$3\kms \citep{yildiz2013} with our $^{13}$CO detection, likely containing the infalling envelope around the protostar (see Section~\ref{sec:13co}). The blueshifted wing of the $^{12}$CO spectrum extends from $<-4$\kms ~to $-71$\kms, and the redshifted component also has high-velocity emission (7.5--55\kms).
The blueshifted components have arc-like structures with a size of several thousand astronomical units, and they surround the continuum source and have a stripe pattern.
In particular, the arc-like structure close to the continuum source resembles a crescent moon.
The redshifted components also show a similar stripe pattern.

To investigate whether the crescent-shaped structure is an interferometric artifact, i.e., due to the missing flux, we compared our data with the data from a single-dish telescope. Using the 15\,meter Swedish-ESO Submillimeter Telescope, \citet{knee1992} showed that the $^{12}$CO\,($J$ = 1--0) high-velocity emission is less than $\sim 0.1$\,K at a velocity of $-20$\,km\,s$^{-1}$ with an angular resolution of $\sim 40''$.
We measured the CO brightness temperature across the entire field of view ($\sim 40'' \times 40''$) at $-20$\,km\,s$^{-1}$ in our data and confirmed that the brightness temperature in our data is consistent with that in  \citet{knee1992}.
Thus, the CO crescent-shaped structure seen in the high-velocity component is not likely to be an overemphasized feature due to the interferometric effect.
As mentioned in Section~\ref{sec:intro}, recent high-resolution molecular gas observations have reported peculiar structures associated with protostars, such as streamers and dynamical interactions among multiple sources.
They are mainly identified in dense gas tracers, e.g., C$^{18}$O and HCO$^{+}$, and their relative velocities are close to the systemic velocity within a few km\,s$^{-1}$.
However, we identified the stripe pattern composed of the multiple arc-like structures in the low-density tracer of $^{12}$CO, and the velocities are too high to be explained by accretion motion.
We regard these components as outflow motions, as already suggested in the previous single-dish works \cite[e.g.,][]{knee1992}.
The shape deviates significantly from the usually identified conical-type feature and can probably be attributed to the viewing angle effect (i.e., close to the face-on configuration).
The interpretation and further discussion of the outflow will be given in Section~\ref{sec:outflow_model}.

\begin{figure*}
\includegraphics[width=180mm]{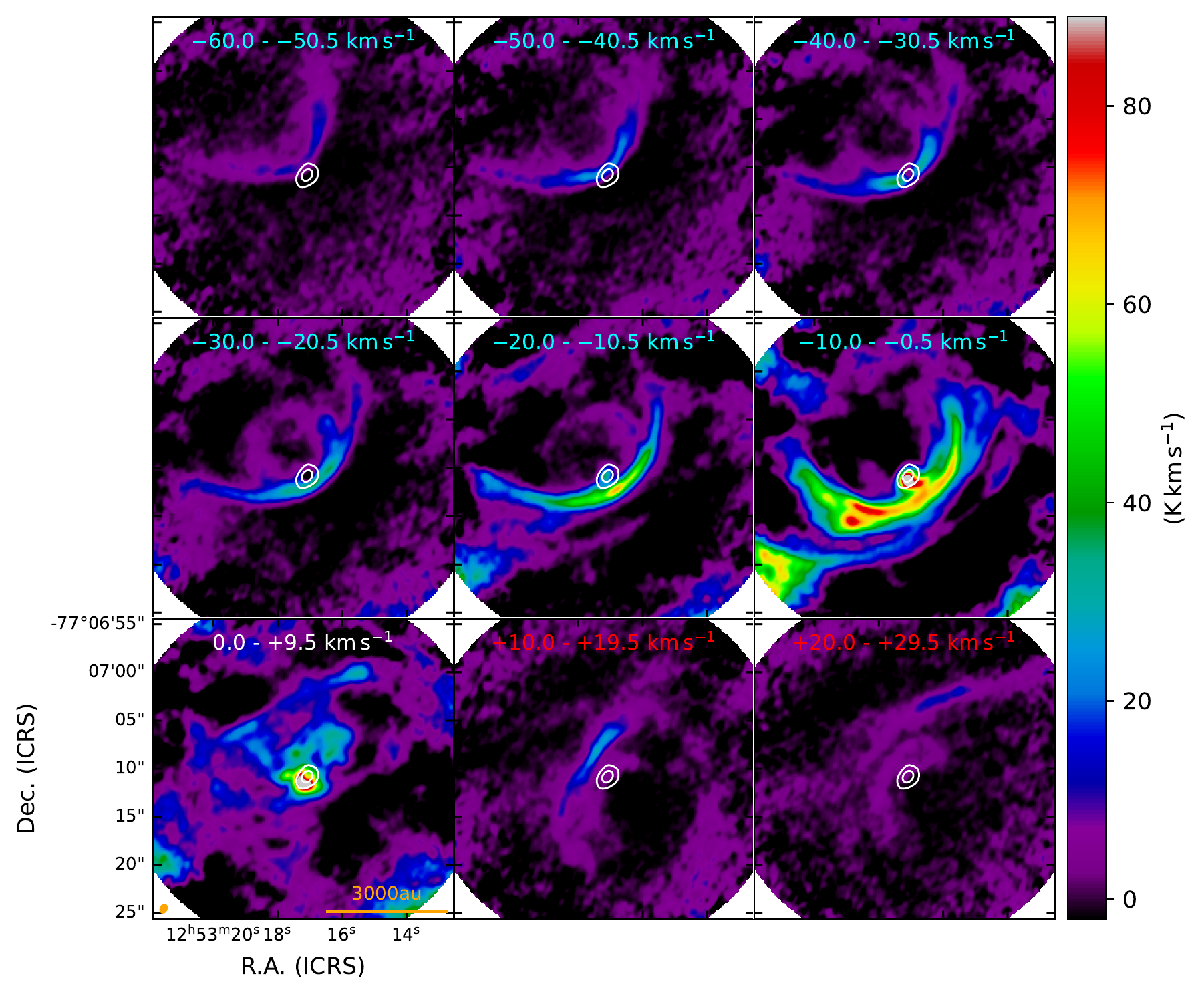}
\caption{
$^{12}$CO velocity-channel maps toward DK Cha.
The color scale shows the integrated intensity for each channel.
The integrated velocity ranges are given in the upper part of each panel; the blue and red text indicate blueshifted and redshifted components, respectively (the white text indicates a velocity range that includes the systemic velocity).
The white contours show the 1.3\,mm continuum emission, and the orange ellipse in the lower left corner represents the synthesized beam size (same as Figure~\ref{fig:rgb}).
\label{fig:12co_channel}}
\end{figure*}

\begin{figure*}
\centering
\includegraphics[width=1.0\columnwidth]{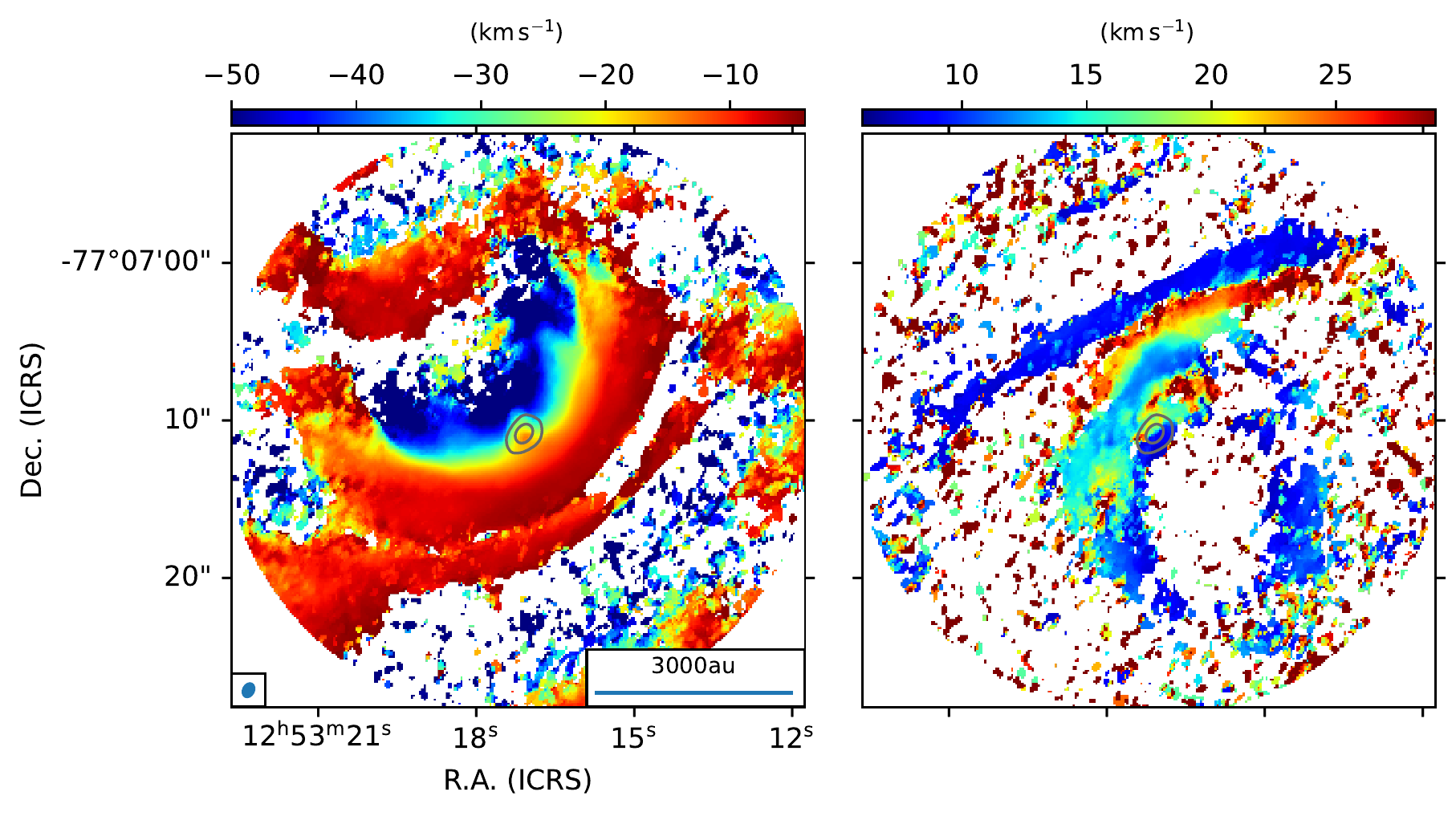}
\caption{
$^{12}$CO moment 1 maps toward DK Cha.
(Left) Moment 1 map of the blueshifted components (color).
The velocity range is $-71$--$-4$\kms.
We used only emissions above $3\sigma$ for each pixel.
The gray contours show the 1.3\,mm continuum emission, and the blue ellipse at the lower left corner represents the synthesized beam size (same as Figure~\ref{fig:rgb}).
(Right) As for the left panel, but for the redshifted components  ($7.5$--$55$\kms).
\label{fig:12co_mom1}}
\end{figure*}

Figure \ref{fig:12co_channel} shows velocity-channel maps of the $^{12}$CO emission in the range $-60$--30\kms.
The range of the velocity integration is 10\,km\,s$^{-1}$ for each channel.
The blueshifted high-velocity component ($\lesssim -40$\,km\,s$^{-1}$) has a crescent-shaped structure and is distributed slightly northeast of the continuum source.
The blueshifted low-velocity component ($-20 \lesssim v_{\rm LSR} \lesssim 0$\,km\,s$^{-1}$) has a similar shape but surrounds the continuum source, and the emitting region seems to be more extended than the high-velocity component.
The redshifted component ($\gtrsim 10$\,km\,s$^{-1}$) has a similar arc-like structure to the blueshifted one, with the opposite orientation.

To verify the characteristics of the outflow described above in more detail, we generated $^{12}$CO moment 1 maps for the blueshifted and redshifted components (Figure~\ref{fig:12co_mom1}).
The figure clearly shows the two-dimensional velocity structure/gradient of the outflow, especially for the blueshifted component.
A blueshifted compact high-velocity emission is located on the northeast side of the continuum source, while a more extended low-velocity emission is distributed on the opposite side, forming a layered velocity gradient across the crescent shape as a whole.

One of the densest regions ($A_V \sim 20$\,mag and $N_{\rm H_2} > 10^{22}$\,cm$^{-2}$) in the Chamaeleon~II cloud is located in the vicinity of DK~Cha \citep[e.g.,][]{alcala2008,alves_de_oliveira2014}.
Compared to the blueshifted outflow, the velocity gradient of the redshifted outflow is not clear.  
In addition, the redshifted outflow velocity is slower than the  blueshifted outflow velocity.
Thus, DK~Cha seems to be in front of the dense region, and the redshifted components could be decelerated due to the interaction with the surrounding dense gas.

\subsection{$^{13}$CO Structures} \label{sec:13co}
\begin{figure*}
\centering
\includegraphics[width=0.9\columnwidth]{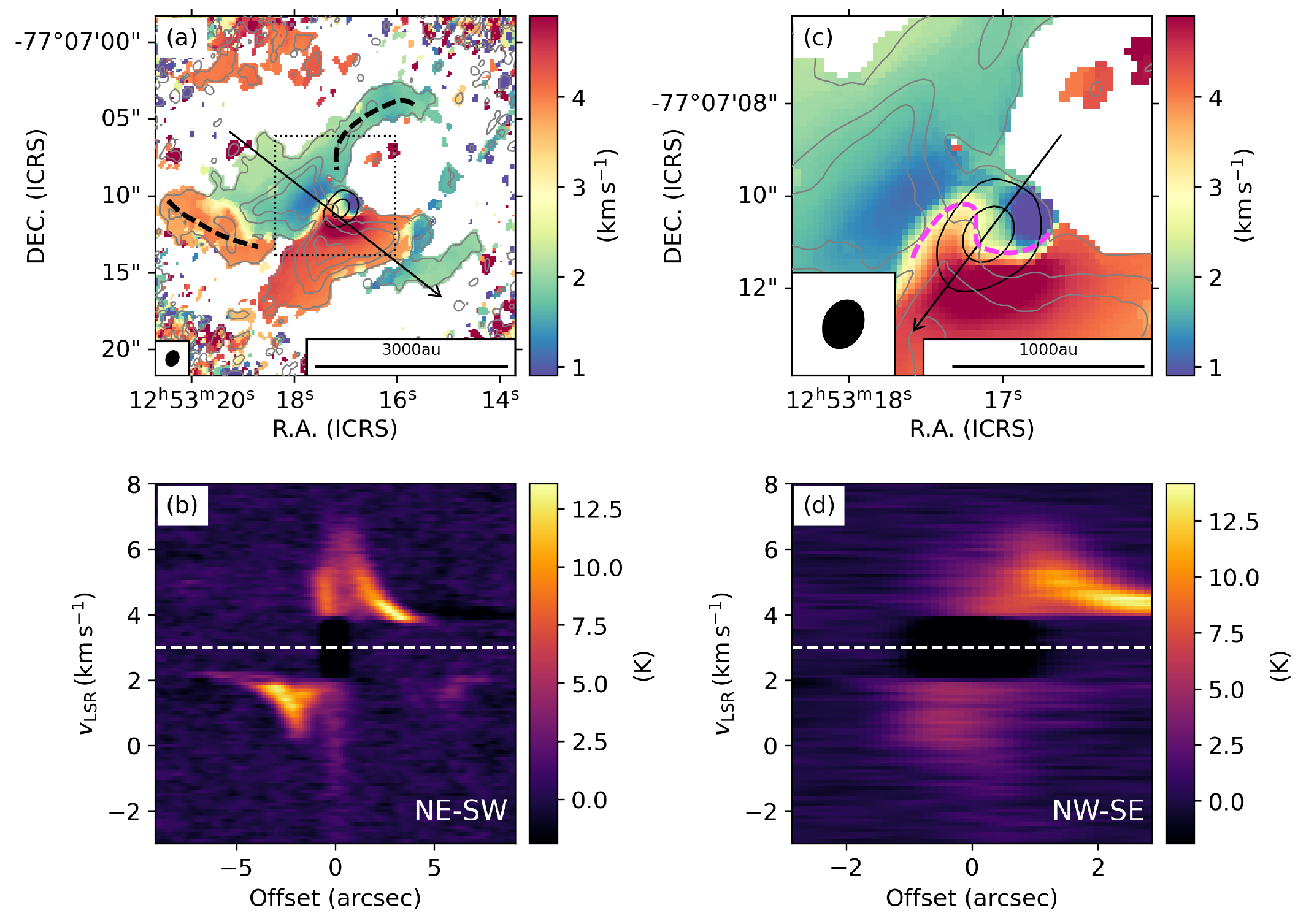}
\caption{
Spatial and velocity structures of $^{13}$CO emission around DK Cha observed by ALMA.
(a)
Moment 1 map of the $>5\sigma$ emission (color).
The gray contours show the $^{13}$CO peak temperature map with 10, 25, 40, and 55$\sigma$, where 1$\sigma$ is 0.26\,K.
The black contours show the 1.3\,mm continuum emission, and the black ellipse at the bottom left corner is the beam size (same as Figure~\ref{fig:rgb}).
The black arrow indicates the direction of the PV diagram shown in panel (b).
(b)
PV diagram along the direction from northeast to southwest.
The white dashed line indicates a velocity of 3\kms, which is the systemic velocity for DK~Cha \citep[e.g.,][]{yildiz2013}.
The position with offset $= 0$ corresponds to the position of the protostar.
(c)
As for panel (a), but zoomed toward the continuum source.
The black arrow indicates the direction of the PV diagram shown in panel (d).
(d)
As for panel (b), but for the direction from northwest to southeast.
\label{fig:13co}}
\end{figure*}

Figure~\ref{fig:13co}~(a) plots the $^{13}$CO\,($J$ = 2--1) moment 1 and the overlaid peak temperature with the gray contours.
The $^{13}$CO image shows an envelope structure with a spatial extent of a few thousand au.
The strong emission on the northeast side is blueshifted ($\sim 1.5$\kms) and that on the southwest side is redshifted ($\sim 4.5$\kms).
This direction is similar to the expected outflow axis.
Figure~\ref{fig:13co}~(b) shows a position--velocity (PV) diagram along that direction; the offset $=0$ point corresponds to the position of the protostar.
The relative velocity tends to be higher in regions closer to the central protostar within $\pm 5''$ ($\sim 1200$\,au), indicating an infalling motion of the envelope gas (for details, see Section~\ref{sec:gas_motion}).

The black dashed lines in Figure~\ref{fig:13co}~(a) show that the $^{13}$CO emission exhibits arc-like structures.
The blueshifted ($\sim2$\kms) and redshifted ($\sim4$\kms) arcs extend to the northwest and northeast toward the continuum source and connect to the infalling envelope.
Both have a length of $\sim1000$\,au.
To further characterize the structures, we estimated their density and mass.
Assuming local thermodynamic equilibrium ($T=20$\,K), we can determine the $^{13}$CO column density using the following equation \citep{wilson2013},
\begin{equation}
    N_{\rm ^{13}CO} \approx 1.5\times10^{14}\frac{T ~e^{5.3/T} ~\int \tau_{\nu} \,{\rm d}v}{1-e^{-10.6/T}} ~{\rm cm^{-2}} ~.
\end{equation}
We obtain $N_{\rm ^{13}CO} = 1.6\times10^{15} \,{\rm cm^{-2}}$.
Assuming that the abundance ratio of $^{12}$CO to $^{13}$CO is 70 and the ratio of H$_2$ to $^{12}$CO is $10^4$ \citep{frerking1982}, we get a H$_2$ column density of $N_{\rm H_2} = 1.1\times10^{21}\,{\rm cm^{-2}}$ and arc mass of $M_{\rm H_2, arc} = 5.7\times10^{-4}\,M_\odot$.
We describe the interpretation of the arc-like structures in Section~\ref{sec:gas_motion}.

Figure~\ref{fig:13co}~(c) is a zoomed-in view of the vicinity of the continuum source, as indicated by the black rectangle in panel (a).
As indicated by the magenta dashed line, the $^{13}$CO emissions of velocities close to the systemic velocity have an S-shaped configuration.
For the continuum source, the velocity gradient extends from northwest to southeast and its extent is about 500\,au.
The direction of the velocity gradient corresponds to the major axis of the disk and is also perpendicular to the outflow axis and to the large-scale velocity gradient seen in panel (a).
Figure~\ref{fig:13co}~(d) shows a PV diagram along the arrow shown in panel (c).
There are emissions with relative velocities of $\gtrsim 3$\,km\,s$^{-1}$ in the region within $\pm 2''$ (corresponding to $\sim 500$\,au) of the continuum source.
The region with a negative offset (i.e., northwest of the disk) is blueshifted and the opposite side is redshifted.
The small-scale velocity gradient perpendicular to the outflow could trace the rotating motion of the disk or the envelope (see also Section~\ref{sec:gas_motion}).

The velocity structure provides information on the protostellar mass, although the angular resolution is not sufficiently high to determine it properly.
At an offset of $\pm$1\arcsec ($\sim$240\,au), the relative velocity is $\sim$2\,km\,s$^{-1}$ (see Figure~\ref{fig:13co}).
To produce such a high velocity at large radii by Keplerian rotation requires at least a 1\,$M_{\odot}$ source with a mass heavier than typical low-mass protostars.
This estimation gives a lower limit of the protostellar mass without considering the inclination angle, which is about a factor of 2 heavier in the case of close to face-on viewing.
The dynamically inferred mass is consistent with the mass estimated by \citet{spezzi2008} based on the position of DK~Cha in the HR diagram.

\section{Discussion} \label{sec:discussion}

\subsection{Interpretation of the Outflow Velocity Structure} \label{sec:outflow_model}

Numerous observational studies have reported outflows, most of which are nicely shaped bipolar flows when viewed nearly edge-on.
It is difficult to spatially resolve the radial velocity distribution within an outflow in such objects because the observed outflow velocity decreases as the line-of-sight direction approaches edge-on. 
In addition, it is also difficult to investigate the internal velocity structure of an outflow when viewed close to edge-on.
We observed the DK~Cha outflow from nearly pole-on and could clearly detect its spatial velocity structure, as shown in the Figures~\ref{fig:rgb}--\ref{fig:12co_mom1}.
In this section, we discuss the three-dimensional velocity structure of the DK~Cha outflow based on the observed velocity gradient in the plane of the sky.

The spatial distribution of the $^{12}$CO emission seen in Figures~\ref{fig:rgb}--\ref{fig:12co_mom1} is highly complicated.
Thus, it is difficult to guess the origin of the emission from the morphology alone.
However, as mentioned in Section~\ref{sec:12co}, their typical velocities are much faster than the speed of sound at 10\,K ($\sim 0.2$\kms), and their complex structures can only be attributed to outflow origin.
Since it is significantly difficult to reproduce blueshifted and redshifted emissions surrounding the continuum source in a typical edge-on view of the outflow, we interpreted the CO emission as a pole-on view of the outflow.
Note that previous studies already suggested that the DK~Cha outflow is observed nearly pole-on \citep{van_kempen2009, yildiz2015}.

\subsubsection{Comparison with Another Pole-on Outflow}
Before discussing the outflow structure of DK~Cha, we compare the DK~Cha outflow with another pole-on outflow.
\citet{fernandez_lopez2020} reported that the Class~II star DO~Tau has a similar outflow structure in the same $^{12}$CO\,($J$ = 2--1) emission line as our detected outflow structure.
They identified ring-shaped outflows with three different radii and velocities.
Based on the disk inclination angle ($i=19^\circ$) and other factors, they interpreted these structures to be a pole-on view of the outflow.
The ring-like structures of DO~Tau are closer to a perfect circle than the crescent-shaped structures of DK~Cha.
This difference may be attributed to the inclination angle, and we expect that the inclination angle of DO~Tau is smaller than that of DK~Cha.
The radii of the rings of DO~Tau range from 220 to 800\,au, slightly smaller than for DK~Cha.
The velocities of the DO~Tau outflow are $-10$--8\,km\,s$^{-1}$.
DK~Cha is considered to be younger than DO~Tau.
The physical properties of DK~Cha, such as the outflow emission, velocity, and spatial scale, are greater than those of DO~Tau.
Thus, DK~Cha could be a more suitable object for observing outflow-driven protostellar systems from nearly face-on directions and for discussing the spatial velocity structure of the outflow.

\subsubsection{Three-dimensional Outflow Structure}
In the following, we discuss which model could reasonably explain the spatial velocity structure of the outflow observed in the CO emission.
As described in Section~\ref{sec:12co}, the redshifted component is likely to interact with the surrounding gas, resulting in a complicated structure.
On the other hand, the blueshifted component has a relatively simple velocity structure.
Thus, we focused only on the velocity structure of the blueshifted outflow (see the left panel of Figure~\ref{fig:12co_mom1}).

\begin{figure*}
\centering
\includegraphics[width=0.8\columnwidth]{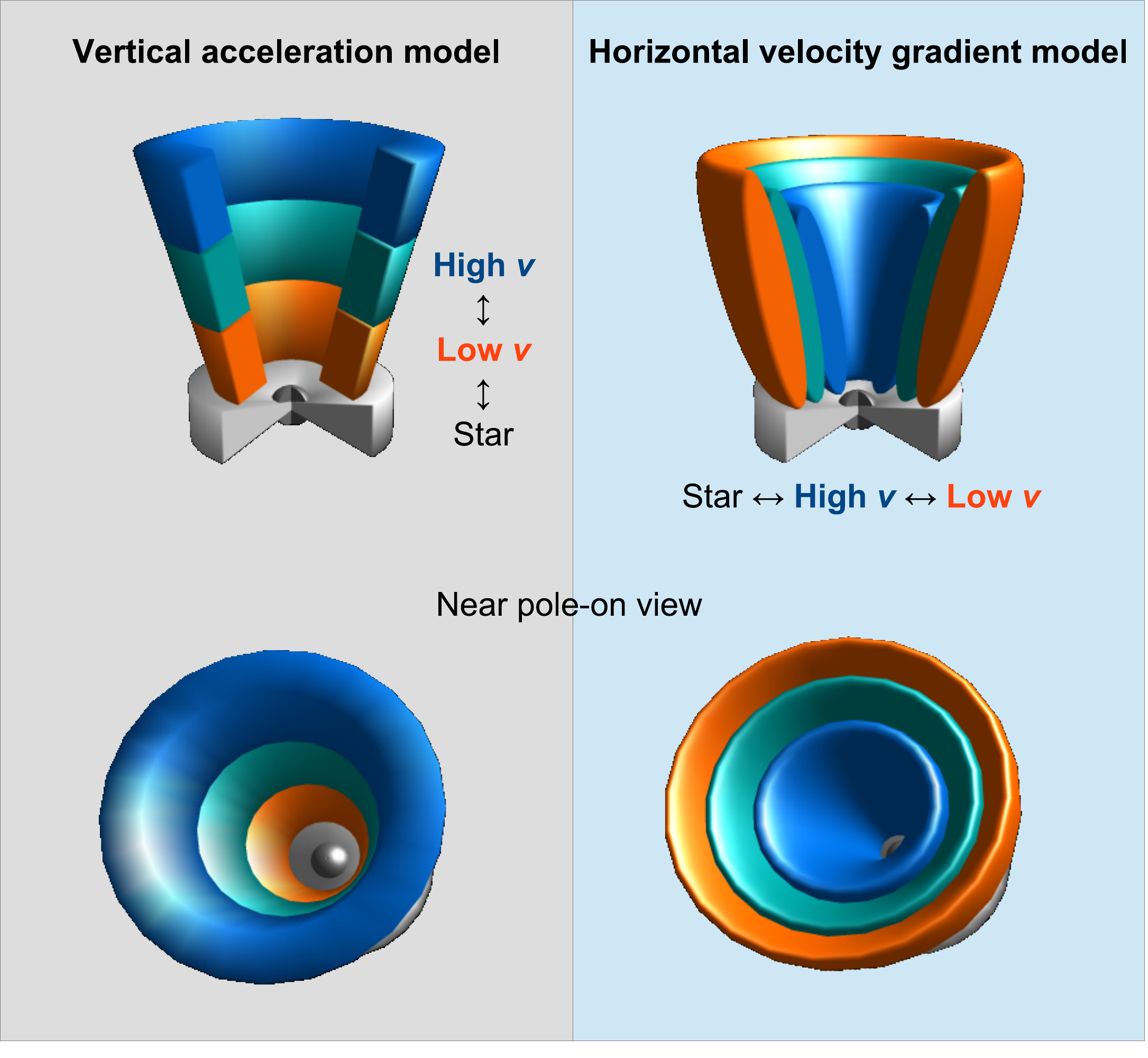}
\caption{
Schematic view of the outflow driven from the protostar.
(Left) Model of an outflow gradually accelerating along the outflow axis.
The orange-, green-, and blue-filled regions have low-, intermediate-, and high-velocity components, respectively.
For simplicity, only three different velocities are shown in the figure.
The dark gray sphere and gray region represent the central protostar and circumstellar disk, respectively.
(Right) Model of an outflow with a horizontal velocity gradient.
\label{fig:outflow}}
\end{figure*}

A simple model proposed by \citet{lee2000} is often used to interpret the outflow velocity structure emerging on the PV diagram.
Their model describes the outflow as a parabolic surface with a velocity structure that accelerates along the outflow axis.
The model appears to explain well the outflow velocity structure on the PV diagram along the outflow axis, while it is not obvious that the model can explain the spatial velocity structure in the plane of the sky.
We, therefore, present a simplified three-dimensional illustration that provides an overview of the velocity structure in the left panels in Figure~\ref{fig:outflow}.
We assume continuous acceleration, while for simplicity we represent the outflow as being composed of three different velocities (orange for the low-, green for the intermediate-, and blue for the high-velocity component).
The lower left panel of Figure~\ref{fig:outflow} clearly shows that the high-velocity component has a more extended structure than the low-velocity component in the plane of the sky.
Figures~\ref{fig:12co_channel} and \ref{fig:12co_mom1} show that the low-velocity component surrounds the high-velocity component, which is the opposite to what is observed for the model reported by \citet{lee2000}.
Thus, it is difficult to reproduce our results with the vertical acceleration model.
Note that we do not mean that outflows are not actually accelerating along the outflow axis.

Another possible approach is to examine the horizontal velocity structure of the outflow (or the velocity distribution in the plane perpendicular to the outflow axis, not along the outflow surface).
We consider the horizontal velocity gradient model, in which the gas in the inner regions flows outward at higher velocity (Figure~\ref{fig:outflow} right panels).
If we observe such an outflow from a direction close to pole-on, a layered velocity structure should appear, as shown in the lower right panel of Figure~\ref{fig:outflow}.
In addition, the high-velocity component is expected to be more compact than the low-velocity one.
The velocity structure of the observed outflow in the left panel of Figure~\ref{fig:12co_mom1} has a similar layered velocity gradient, with a compact high-velocity component surrounded by an extended low-velocity component.
In the panel, the missing part of the ellipse could be due to interactions with the ambient gas, lack of sensitivity outside the observation area, or other factors.
Therefore, our observations reveal that outflows with different velocities are driven from different radii.
Such a velocity structure can be naturally reproduced by an outflow directly driven by the magnetocentrifugal mechanism \citep[e.g.,][]{machida2014,machida2019}.

\subsection{Outflow Launching Regions}
The outflow driving mechanism is still under debate.
There are two main scenarios.
In one, the outflow is produced by ambient gas receiving momentum from the high-velocity jet \citep[e.g.,][]{arce2007}.
On the other, the outflow is driven directly from the circumstellar disk by the magnetocentrifugal mechanism irrespective of the appearance of the jet \citep[e.g.,][]{machida2014}.
Recent ALMA observations have supported the latter.
For example, \citet{bjerkeli2016} reported that the TMC-1A outflow is driven from the outer edge of the disk.
A clear signature of rotation could be detected in the bipolar outflow of Orion Source I \citep{hirota2017}. The large angular momentum of the outflow indicates that the outflow is driven far from the central protostar.
Note that the outflow driven from the region close to the central star cannot have a large angular momentum (or rapid rotation).
\citet{matsushita2019} discovered an extremely high-velocity flow and outflow around MMS~5 in the Orion Molecular Cloud~3, and found that their axes are misaligned, whereas if the low-velocity flow is entrained by the high-velocity flow, their axes should be aligned.

Here, based on the directly driven outflow scenario, we estimate the driving radius from the observed outflow velocities shown in Figures~\ref{fig:rgb}--\ref{fig:12co_mom1}.
Classical theoretical studies \citep[e.g.,][]{mestel1961,mestel1968} have shown that the terminal velocity of the outflow (or disk wind) roughly corresponds to the Keplerian velocity at its launching radius.
Thus, we simply assume that the observed outflow velocity $v_{\rm out}$ agrees with the Keplerian velocity $v_{\rm K}$ at each launching radius $r_0$,
\begin{equation}
    v_{\rm out} \sim v_{\rm K} = \sqrt{\frac{G M_{\rm star}}{r_0}}.
    \label{eq:vout}
\end{equation}
With equation~(\ref{eq:vout}), the outflow launching radius $r_0$ can be described as
\begin{equation}
    r_0 \sim G M_{\rm star}\, v^{-2}_{\rm out}
    = 0.7 \,{\rm au} \left( \frac{M_{\rm star}}{2\msun} \right)
        \left( \frac{v_{\rm out}}{50\,\rm km\,s^{-1}} \right)^{-2},
\label{eq:r0}
\end{equation}
where we have used a protostellar mass $M_{\rm star}$ of $2\msun$ \citep{spezzi2008}.
As shown in Figures~\ref{fig:rgb}--\ref{fig:12co_mom1}, the outflow of DK~Cha has crescent-shaped structures across the low-velocity ($< 5$\kms) to the high-velocity ($> 50$\kms) components.
The outflow launching region extends over two orders of magnitude ($\sim 0.7$--70\,au) if these outflow components with different velocities are driven directly from their respective radii.
Note that the driving radius depends somewhat on the Alfvén radius (i.e., the intensity of the magnetic field and ram pressure).
Thus, there is uncertainty in estimating the launching radius within a factor of $\lesssim 10$
   \citep[e.g.,][]{pudritz1986,koenigl1993,kudoh1997ApJ,pudritz2019}.
Therefore, the inner ($\sim0.7$\,au) and outer ($\sim 70$\,au) launching radii can be changed according to the extent of the Alfvén radius.
However, the large extent of the outflow driving region (over two orders of magnitude in radius) would not change significantly, because the outflow velocity at each shell (see, Figure~\ref{fig:outflow} right panels) is primarily determined by the Keplerian velocity at each launching radius \citep{kudoh1997ApJ,Kudoh1997b,Machida2008} and the difference in the outflow velocity ($\sim5-50$\kms) gives the difference in the launching radius proportional to the outflow velocity (see, eq.~[\ref{eq:r0}]).

Considering the conservation of the (specific) angular momentum ($j = r_{\rm out} v_{\rm rot} = \sqrt{G M_{\rm star} r_0}$), it is valuable to estimate the rotational velocity of the outflow $v_{\rm rot}$,
\begin{equation}
  v_{\rm rot} = 0.4\,{\rm km\,s^{-1}} \left( \frac{r_{\rm out}}{1000\,\rm au} \right)^{-1}
    \left( \frac{M_{\rm star}}{2\msun} \right)^{0.5} \left( \frac{r_0}{70\,\rm au} \right)^{0.5},
\end{equation}
where $r_{\rm out}$ is the observed outflow radius (the half width of the outflow lobe).
Thus, the rotational velocity of the outflow is clearly less than the velocity resolution in this observation.
Since the outflow rotational velocity becomes smaller due to the inclination angle, it is natural that we could not detect any sign of outflow rotation.

\subsection{Rotating Infalling Envelope and Streamer} \label{sec:gas_motion}

In this subsection, we discuss the gas motion expected from the $^{13}$CO\,($J$ = 2--1) emission.
As described in Section~\ref{sec:13co}, two velocity gradients around DK~Cha are detected in the $^{13}$CO emission (Figure~\ref{fig:13co}), one perpendicular and one parallel to the outflow axis.
Relative velocities seem to be higher in regions closer to the central protostar.
This velocity structure is similar to the infalling and rotating gas motion reported in recent ALMA observations \citep[e.g.,][]{aso2015,sakai2016,yen2017}.
In particular, the velocity structure (moment 1 map) of HL~Tau \citep[][Figure~1]{yen2017} is very similar to our results (Figure~\ref{fig:13co}~(a))
in that, both sources have a 1000\,au-scale envelope gas, velocity gradients parallel and perpendicular to the outflow axis, and an S-shaped structure with the gas having velocities close to the systemic velocity in the $^{13}$CO\,($J$ = 2--1) emission.
Thus, it seems that the $^{13}$CO emission has traced the rotating infalling envelope.

We can also confirm the existence of $^{13}$CO arc-like structures in Figure~\ref{fig:13co}~(a).
There is no sign of a binary star in DK~Cha and thus the structure is unlikely to be the result of gravitational interactions in a multiple system \citep{matsumoto2015} like MC27/L1521F \citep{tokuda2014,tokuda2016}.
Rather, it is more natural to interpret these structures as non-axisymmetric accretion flows toward the central protostellar system since these are connected to the rotating infalling envelope described above.
In other words, we expect that the structures are the same phenomenon as the streamers recently observed \citep[][and references therein]{pineda2022}.
Among Class~I objects, streamers have been detected in HL~Tau and Per-emb-50 with lengths of 520 and 3000\,au, respectively \citep{yen2019,valdivia2022}.
The streamers detected in DK~Cha have a similar size ($\sim$1000\,au) to these objects.
Our observations may be a case in which the rotating infalling envelope and the streamers were detected simultaneously.
HL~Tau, Per-emb-50, and DK~Cha are all intermediate-mass protostars ($\sim 2\msun$). Therefore, it is interesting to investigate whether streamers are associated with a broader range of protostellar mass.
In addition, since shock tracers such as SO have been detected together with streamers \citep{garufi2022}, it will be important to observe different molecular emission lines for DK~Cha.

\section{Conclusions} \label{sec:conclusions}
We have analyzed the ALMA archival data toward the intermediate-mass protostar DK~Cha in the Chamaeleon~II star-forming region.
Our results are summarized as follows.
\begin{enumerate}
    \item A $^{12}$CO\,($J$ = 2--1) emission was detected in the very high-velocity outflow ($|v_{\rm LSR}-v_{\rm sys}| \ge 70$\kms) driven by DK Cha, which exhibits a $\sim$3000\,au-scale crescent-shaped structure surrounding the 1.3\,mm continuum source.
    The outflow has a layered velocity gradient; the lower-velocity component is located southwest toward the continuum peak, and the higher-velocity component is shifted northeastward.
    These observations revealed the spatially resolved velocity structure of the pole-on outflow.
    \item We considered two models to explain the velocity structure of the CO outflow revealed in this study.
    One is to interpret the observed velocity structure as a vertical acceleration of the outflow.
    The other interprets the observed velocity structure as a horizontal velocity gradient of the outflow.
    Based on the fact that the high-velocity component is more compact than the low-velocity one, the latter model is more plausible to explain the velocity gradient of the outflow observed in the CO emission.
    Furthermore, based on the directly driven outflow scenario, we clarified the outflow driving region from the observed outflow velocities.
    The results suggest that the driving region extends over two orders of magnitude ($\sim 0.7$--70\,au).
    \item We detected two different velocity gradients in the $^{13}$CO\,($J$ = 2--1) emission.
    One is a 500\,au-scale compact velocity gradient around the protostar from northwest to southeast, which is perpendicular to the expected axis of the $^{12}$CO outflow.
    The other is a 2500\,au-scale velocity gradient extending from northeast to southwest, parallel to the expected outflow axis.
    The relative velocity $|v_{\rm LSR}-v_{\rm sys}|$ is $> 3$\kms and tends to be higher in regions closer to the central protostar.
    Considering the directions of the gradients and the acceleration toward the protostar, the $^{13}$CO emission traces the rotating infalling envelope.
    In addition, the $^{13}$CO emission shows two arc-like structures.
    One has a blueshifted central velocity of $\sim 2$\kms and extends to the northwest toward DK~Cha.
    The other has a redshifted central velocity of $\sim 4$\kms and extends to the northeast.
    Both have a length of $\sim 1000$\,au.
    Since these structures are connected to the infalling envelope, we concluded that the arc-like structures correspond to non-axisymmetric accretion flows/streamers, often reported in recent years.
\end{enumerate}
DK~Cha has the potential to become a reasonable template for understanding the characteristics of face-on protostellar systems, as it allows us to simultaneously observe various phenomena, such as a crescent-shaped outflow, a rotating infalling envelope, and streamers.
It is expected that future wide-field and deep observations of DK~Cha will help us to further understand the inner and outer structures in more detail.

\begin{acknowledgments}
We would like to thank the anonymous referees for their useful comments that we applied to improve our manuscript. This paper makes use of the following ALMA data: ADS/JAO. ALMA\#2019.1.01792.S. ALMA is a partnership of ESO (representing its member states), the NSF (USA), and NINS (Japan), together with the NRC (Canada), MOST, ASIAA (Taiwan), and KASI (Republic of Korea), in cooperation with the Republic of Chile. The Joint ALMA Observatory is operated by the ESO, AUI/NRAO, and NAOJ. This work was supported by a NAOJ ALMA Scientific Research grant (No. 2022-22B), Grants-in-Aid for Scientific Research (KAKENHI) of the Japan Society for the Promotion of Science (JSPS; grant Nos. JP17H06360, JP17K05387, JP17KK0096, JP21K13962, JP21H00049, JP21K03617, and JP21H00046).
\end{acknowledgments}

\vspace{5mm}
\software{CASA \citep[v5.6.1-8;][]{casa2022}, Astropy \citep{astropy2018}}

\bibliography{references}{}
\bibliographystyle{aasjournal}

\end{document}